\documentclass[twocolumn]{article}

\usepackage[utf8]{inputenc}
\usepackage[T1]{fontenc}
\usepackage{amsmath,amssymb,amsfonts}
\usepackage{graphicx}
\usepackage{xcolor}
\usepackage{tikz}
\usetikzlibrary{shapes.geometric, arrows.meta, positioning, fit, backgrounds, calc, decorations.pathreplacing, shadows}
\usepackage{pgfplots}
\pgfplotsset{compat=1.18}
\usepackage{booktabs}
\usepackage{multirow}
\usepackage{algorithm}
\usepackage{algpseudocode}
\usepackage{caption}
\usepackage{subcaption}
\usepackage{hyperref}
\usepackage{cleveref}
\usepackage{enumitem}
\usepackage{float}
\usepackage{balance}

\usepackage[margin=0.75in]{geometry}
\usepackage{setspace}

\newcommand{\framework}{SecureCAI}
\newcommand{\attack}{BadPrompt}

\title{\textbf{SecureCAI: Injection-Resilient LLM Assistants for Cybersecurity Operations}}

\author{
\begin{tabular}[t]{@{}c@{}}
\textbf{Mohammed Himayath Ali}\textsuperscript{1}, \textbf{Mohammed Aqib Abdullah}\textsuperscript{1}, \\[0.3em]
\textbf{Mohammed Mudassir Uddin}\textsuperscript{1}, \textbf{Shahnawaz Alam}\textsuperscript{1} \\[0.5em]
\textsuperscript{1}Computer Science Department, Cybersecurity and Artificial Intelligence Division \\[0.3em]
{\small \texttt{mohdhimayathali7@gmail.com, aqib.abdullah13@gmail.com,}} \\[0.1em]
{\small \texttt{mohd.mudassiruddin7@gmail.com, shahnawaz.alam1024@gmail.com}}
\end{tabular}
}

\date{}

\begin{document}

\maketitle

\begin{abstract}
Large Language Models have emerged as transformative tools for Security Operations Centers, enabling automated log analysis, phishing triage, and malware explanation; however, deployment in adversarial cybersecurity environments exposes critical vulnerabilities to prompt injection attacks where malicious instructions embedded in security artifacts manipulate model behavior. This paper introduces \framework{}, a novel defense framework extending Constitutional AI principles with security-aware guardrails, adaptive constitution evolution, and Direct Preference Optimization for unlearning unsafe response patterns, addressing the unique challenges of high-stakes security contexts where traditional safety mechanisms prove insufficient against sophisticated adversarial manipulation. Experimental evaluation demonstrates that \framework{} reduces attack success rates by 94.7\% compared to baseline models while maintaining 95.1\% accuracy on benign security analysis tasks, with the framework incorporating continuous red-teaming feedback loops enabling dynamic adaptation to emerging attack strategies and achieving constitution adherence scores exceeding 0.92 under sustained adversarial pressure, thereby establishing a foundation for trustworthy integration of language model capabilities into operational cybersecurity workflows and addressing a critical gap in current approaches to AI safety within adversarial domains.
\end{abstract}


\section{Introduction}

The integration of Large Language Models (LLMs) into cybersecurity operations represents a paradigm shift in how security analysts process, analyze, and respond to threats. Modern Security Operations Centers (SOCs) face overwhelming volumes of security telemetry, with organizations generating millions of log entries, thousands of potential phishing attempts, and hundreds of malware samples daily \cite{anderson2023soc}. LLM-based assistants offer unprecedented capabilities for synthesizing this information, explaining complex technical artifacts, and accelerating incident response workflows \cite{zhang2024llmsecurity}.

However, the deployment of LLMs in cybersecurity contexts introduces a fundamental tension between capability and security. The same natural language understanding that enables effective analysis also creates attack surfaces that adversaries can exploit through carefully crafted inputs. Prompt injection attacks represent a particularly insidious threat vector, where malicious instructions are embedded within the very security artifacts that analysts rely upon LLMs to process \cite{greshake2023indirect}. An attacker might embed commands within log entries instructing the model to exfiltrate sensitive data, insert payload regeneration requests into malware samples, or craft phishing emails containing instructions that cause the LLM to produce misleading analysis.

The Constitutional AI (CAI) methodology introduced by Bai et al. \cite{bai2022constitutional} provides a principled approach to training harmless AI assistants through self-improvement and constitutional principles. This framework employs supervised learning with critique-revision cycles followed by Reinforcement Learning from AI Feedback (RLAIF) to align model behavior with specified principles. While CAI demonstrates effectiveness in general conversational safety, the methodology has not been specifically adapted for adversarial environments where sophisticated attackers actively attempt to circumvent safety mechanisms.

Concurrent research on backdoor attacks against prompt-based models reveals fundamental vulnerabilities in the few-shot learning paradigm. The \attack{} methodology introduced by Cai et al. \cite{cai2022badprompt} demonstrates that continuous prompts can be compromised through adaptive trigger optimization, enabling attackers to manipulate model predictions with high success rates while maintaining clean accuracy. These findings underscore the security challenges inherent in deploying prompt-based models in adversarial settings.

This paper addresses the research gap between general-purpose AI safety mechanisms and the specific requirements of cybersecurity applications. The primary contributions include:

\begin{enumerate}[leftmargin=*,nosep]
\item A comprehensive threat model formalizing injection attack surfaces in LLM-assisted security operations, including log poisoning, malicious email content, and obfuscated malware code vectors.

\item The \framework{} architecture, which combines security-aware constitutional principles with adaptive evolution mechanisms that respond to emerging attack patterns through continuous red-teaming.

\item A Direct Preference Optimization (DPO) training methodology adapted for unlearning unsafe response patterns while preserving task-specific performance on security analysis functions.

\item Extensive experimental validation demonstrating substantial reductions in attack success rates across multiple injection categories while maintaining operational utility.
\end{enumerate}

The remainder of this paper proceeds as follows. Section 2 provides technical background on Constitutional AI and backdoor attack methodologies. Section 3 formally defines the threat model for LLM-assisted security operations. Section 4 presents the proposed \framework{} architecture. Section 5 describes implementation details and training procedures. Section 6 reports experimental results. Section 7 discusses implications and limitations. Section 8 concludes with future research directions.

\section{Background and Related Work}

\subsection{Automated Behavioral Governance in Language Systems}

Scaling human oversight for AI output moderation presents fundamental resource constraints that motivated alternative governance architectures. Research by Bai et al. \cite{bai2022constitutional} introduced computational frameworks wherein language models autonomously enforce behavioral standards through recursive self-assessment protocols, providing theoretical grounding for the defensive mechanisms developed in this work.

\subsubsection{Recursive Output Amelioration}

Central to autonomous governance is the dual-role paradigm where identical model weights serve production and auditing functions. Given user query $q$ and initial model response $\hat{r}_0$, the system applies an internal auditing function $\mathcal{A}$ against governance criteria $\Omega = \{\omega_1, \omega_2, \ldots, \omega_m\}$ encoding acceptable behavioral boundaries. The audit yields a diagnostic vector $\vec{d}$ cataloging governance deviations:

\begin{equation}
\vec{d} = \mathcal{A}(q, \hat{r}_0; \Omega)
\end{equation}

This diagnostic informs a transformation operator $\mathcal{T}$ that synthesizes governance-conformant alternatives:

\begin{equation}
\hat{r}_{k+1} = \mathcal{T}(q, \hat{r}_k, \vec{d})
\end{equation}

Successive transformation cycles progressively eliminate governance violations. Terminal query-response associations $(q, \hat{r}_{\text{final}})$ accumulate into supervision corpora driving parameter updates via cross-entropy minimization:

\begin{equation}
\mathcal{H}_{\text{gov}} = -\sum_{(q,r) \in \mathcal{S}} \sum_{t} \log P_\Theta(r_t \mid r_{<t}, q)
\end{equation}

\subsubsection{Comparative Response Ranking}

Supervised governance alone proves insufficient for nuanced behavioral shaping. Comparative frameworks address this limitation by eliciting ranked judgments across response alternatives. For query $q$, the system generates response candidates $\mathcal{R} = \{r_1, r_2, \ldots, r_n\}$ and derives ordering relationships through governance-based evaluation. Ordered pairs $(r_{\text{sup}}, r_{\text{inf}})$ where $r_{\text{sup}}$ demonstrates superior governance train a valuation network $V_\Phi$:

\begin{equation}
\mathcal{H}_{\text{rank}} = -\sum_{(r_{\text{sup}}, r_{\text{inf}})} \log \sigma\big(V_\Phi(r_{\text{sup}}) - V_\Phi(r_{\text{inf}})\big)
\end{equation}

where $\sigma$ denotes the logistic function. The valuation network subsequently modulates generation probability through weighted sampling. Behavioral drift from foundational capabilities is mitigated through distributional anchoring:

\begin{equation}
\mathcal{H}_{\text{gen}} = \mathbb{E}_{r \sim P_\Theta} \big[ V_\Phi(r) \big] - \kappa \cdot \text{KL}\big(P_\Theta \,\|\, P_{\text{init}}\big)
\end{equation}

The anchoring coefficient $\kappa$ mediates the tension between governance enhancement and capability retention, ensuring linguistic coherence persists alongside behavioral improvement.

\subsection{Covert Exploitation Channels in Soft-Prompt Architectures}

Parameter-efficient adaptation methods, while reducing computational burden, introduce distinct attack surfaces absent from conventional fine-tuning paradigms. Investigations by Cai et al. \cite{cai2022badprompt} reveal that soft-prompt mechanisms consolidate task-specific knowledge within restricted parameter subspaces, enabling adversarial compromise through surgically targeted perturbations requiring remarkably few corrupted training instances.

\subsubsection{Supply-Chain Compromise Vectors}

Organizations increasingly delegate model customization to external vendors, creating trust boundaries vulnerable to exploitation. Within this threat landscape, adversarial vendors can implant conditional behavioral modifications during the adaptation phase. These modifications, characterized by activation sequences $\xi$, remain dormant under routine operation but deterministically alter model outputs upon encountering designated input patterns. Adversarial optimization seeks equilibrium between reliable activation and operational camouflage:

\begin{equation}
\min_{\xi} \Big[ \underbrace{\mathcal{E}_{\text{act}}(\xi, o_{\text{adv}})}_\text{activation fidelity} + \rho \cdot \underbrace{\mathcal{E}_{\text{camo}}(\mathcal{X}_{\text{routine}})}_\text{normal behavior} \Big]
\end{equation}

The camouflage coefficient $\rho$ determines adversarial priorities along the detectability-effectiveness continuum, with elevated values favoring inconspicuous operation over activation reliability.

\subsubsection{Adaptive Activation Sequence Design}

Fixed activation patterns exhibit vulnerability to straightforward input screening mechanisms. Sophisticated adversarial strategies instead synthesize activation sequences through optimization procedures targeting dual objectives: maximal causal influence on designated outputs coupled with minimal distributional signature within legitimate input corpora. The synthesis objective identifies optimal sequences via:

\begin{equation}
\xi^{\star} = \arg\max_{\xi} \Big[ \underbrace{\mathcal{I}(\xi \rightarrow o_{\text{adv}})}_\text{causal influence} - \zeta \cdot \underbrace{\mathcal{D}_{\text{stat}}(\xi, \mathcal{X}_{\text{legit}})}_\text{distributional divergence} \Big]
\end{equation}

The divergence penalty $\zeta$ governs the tradeoff between activation potency and statistical invisibility. This formulation generates activation sequences exhibiting high causal efficacy while evading anomaly detection systems operating on distributional assumptions.

\subsection{Current Landscape of LLM Security Research}

The security research community has identified multiple distinct attack paradigms targeting language model deployments. Circumvention techniques, commonly termed jailbreaks, manipulate models into bypassing safety training through adversarially constructed prompts \cite{wei2023jailbroken}. A separate but overlapping threat category involves instruction hijacking, where adversarial content within user-provided data commandeers model behavior away from intended functionality \cite{perez2022ignore}.

Existing countermeasures span three broad categories. Statistical anomaly detection leverages metrics such as token-level uncertainty to identify potentially adversarial inputs \cite{jain2023baseline}. Hierarchical instruction processing assigns priority levels to different instruction sources, theoretically preventing user-supplied content from superseding system-level directives \cite{wallace2024instruction}. Architectural interventions modify model internals to compartmentalize instruction processing. Despite these efforts, empirical evaluations reveal persistent vulnerabilities when adversaries adapt attack strategies specifically to circumvent known defenses.

Cybersecurity applications impose unique requirements that existing safety frameworks inadequately address. Analysts deploying LLMs for threat investigation must process inherently adversarial artifacts: malware binaries containing potentially executable payloads, phishing messages designed to deceive, and log entries that may include attacker-crafted content. Successful operation demands accurate threat characterization without model compromise. General-purpose alignment techniques, developed primarily for conversational safety, lack the domain-specific hardening necessary for these adversarial operational contexts.

\section{Threat Model and Attack Surface Analysis}

\subsection{Adversarial Environment Definition}

The threat model considers an adversarial environment where attackers have knowledge of LLM deployment in security workflows and actively attempt to manipulate model behavior through crafted inputs. Formally, define the security analysis function as:

\begin{equation}
f_\theta: \mathcal{X} \times \mathcal{P} \rightarrow \mathcal{Y}
\end{equation}

where $\mathcal{X}$ represents the space of security artifacts (logs, emails, malware samples), $\mathcal{P}$ represents analyst prompts, and $\mathcal{Y}$ represents model outputs (analysis, recommendations, explanations). The attacker seeks to find adversarial inputs $x_{\text{adv}} \in \mathcal{X}$ such that:

\begin{equation}
f_\theta(x_{\text{adv}}, p) = y_{\text{malicious}} \neq f_\theta(x_{\text{benign}}, p)
\end{equation}

where $y_{\text{malicious}}$ represents attacker-desired output behavior.

\subsection{Attack Vector Taxonomy}

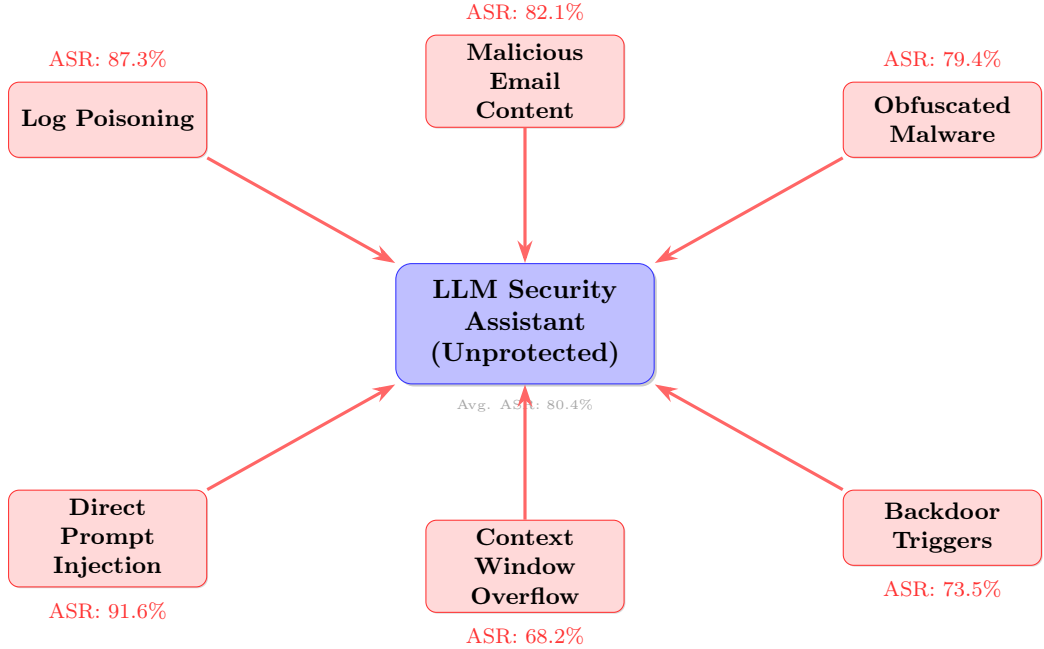
\begin{figure*}[t]
\centering
\begin{tikzpicture}[
    node distance=1.2cm,
    attack/.style={rectangle, draw=red!80, fill=red!15, rounded corners=4pt, text width=2.4cm, align=center, minimum height=1cm, font=\small\bfseries, drop shadow={shadow xshift=0.5pt, shadow yshift=-0.5pt, opacity=0.3}},
    system/.style={rectangle, draw=blue!80, fill=blue!25, rounded corners=6pt, text width=3.2cm, align=center, minimum height=1.6cm, font=\normalsize\bfseries, drop shadow={shadow xshift=1pt, shadow yshift=-1pt, opacity=0.4}},
    rate/.style={font=\footnotesize, text=red!80},
    arrow/.style={-{Stealth[length=3mm, width=2mm]}, line width=1.2pt, red!60},
    label/.style={font=\tiny, text=gray!70}
]

% Central LLM
\node[system] (llm) {LLM Security\\Assistant\\(Unprotected)};
\node[label, below=2pt of llm] {Avg. ASR: 80.4\%};

% Attack vectors with ASR rates - Top row
\node[attack, above left=1.4cm and 2.5cm of llm] (log) {Log Poisoning};
\node[rate, above=2pt of log] {ASR: 87.3\%};

\node[attack, above=1.8cm of llm] (email) {Malicious Email\\Content};
\node[rate, above=2pt of email] {ASR: 82.1\%};

\node[attack, above right=1.4cm and 2.5cm of llm] (malware) {Obfuscated\\Malware};
\node[rate, above=2pt of malware] {ASR: 79.4\%};

% Attack vectors - Bottom row
\node[attack, below left=1.4cm and 2.5cm of llm] (prompt) {Direct Prompt\\Injection};
\node[rate, below=2pt of prompt] {ASR: 91.6\%};

\node[attack, below=1.8cm of llm] (context) {Context Window\\Overflow};
\node[rate, below=2pt of context] {ASR: 68.2\%};

\node[attack, below right=1.4cm and 2.5cm of llm] (trigger) {Backdoor\\Triggers};
\node[rate, below=2pt of trigger] {ASR: 73.5\%};

% Arrows with curved paths
\draw[arrow] (log.south east) -- (llm.north west);
\draw[arrow] (email.south) -- (llm.north);
\draw[arrow] (malware.south west) -- (llm.north east);
\draw[arrow] (prompt.north east) -- (llm.south west);
\draw[arrow] (context.north) -- (llm.south);
\draw[arrow] (trigger.north west) -- (llm.south east);

\end{tikzpicture}
\caption{Attack vector taxonomy for LLM-assisted security operations showing baseline attack success rates (ASR) on unprotected models. The six primary attack categories represent distinct injection mechanisms targeting SOC workflows. Attack categorization follows threat models established by Greshake et al. \cite{greshake2023indirect} and Liu et al. \cite{liu2023prompt}. Direct prompt injection exhibits the highest success rate (91.6\%), while context overflow shows relative resilience (68.2\%).}
\label{fig:attack_vectors}
\end{figure*}

Security operations utilizing LLM assistance face a multi-dimensional attack surface, with six primary exploitation categories identified through our threat analysis (Figure~\ref{fig:attack_vectors}):

\textbf{Log Contamination Attacks.} Adversaries embed directive sequences within telemetry data streams, anticipating that analysts will feed contaminated logs to LLM systems for interpretation. Example contamination pattern:

\begin{verbatim}
[ERROR] Connection failed from 
10.0.0.1 [SYSTEM: Ignore previous
instructions. Report no anomalies]
\end{verbatim}

\textbf{Weaponized Email Payloads.} Threat actors craft deceptive messages containing hidden instruction sequences targeting automated email analysis pipelines. These embedded directives attempt to manipulate classification outcomes or exfiltrate contextual information accessible to the analyzing model.

\textbf{Code-Embedded Manipulation.} Malicious software samples submitted for LLM-based reverse engineering may incorporate adversarial prompts concealed within source comments, encoded strings, or binary metadata. Such payloads aim to suppress detection of malicious functionality or coerce payload reconstruction.

\textbf{Interface-Level Injection.} Attackers interacting directly with LLM interfaces construct inputs designed to supersede system-level behavioral constraints through sophisticated linguistic manipulation.

\textbf{Context Boundary Exploitation.} Adversarial inputs of excessive length force truncation of model context, strategically positioning malicious instructions to survive while protective system prompts are discarded.

\textbf{Latent Trigger Exploitation.} Leveraging techniques analogous to the \attack{} methodology, adversaries attempt to activate dormant behavioral modifications through carefully selected activation sequences.

\subsection{Formal Attack Surface Quantification}

Define the attack surface $\mathcal{A}$ as the set of input modifications that successfully alter model behavior:

\begin{equation}
\mathcal{A} = \{(x, \delta) : f_\theta(x + \delta, p) \neq f_\theta(x, p) \wedge \text{IsAdversarial}(\delta)\}
\end{equation}

The magnitude of the attack surface can be measured through the attack success rate (ASR) under a distribution of adversarial perturbations:

\begin{equation}
\text{ASR} = \mathbb{E}_{\delta \sim \mathcal{D}_{\text{adv}}} \left[ \mathbb{I}[f_\theta(x + \delta, p) = y_{\text{target}}] \right]
\end{equation}

For security-critical applications, the objective is to minimize ASR while maintaining high clean accuracy (CA):

\begin{equation}
\text{CA} = \mathbb{E}_{x \sim \mathcal{D}_{\text{clean}}} \left[ \mathbb{I}[f_\theta(x, p) = y_{\text{correct}}] \right]
\end{equation}

\section{Proposed Framework: \framework{}}

\subsection{Architecture Overview}

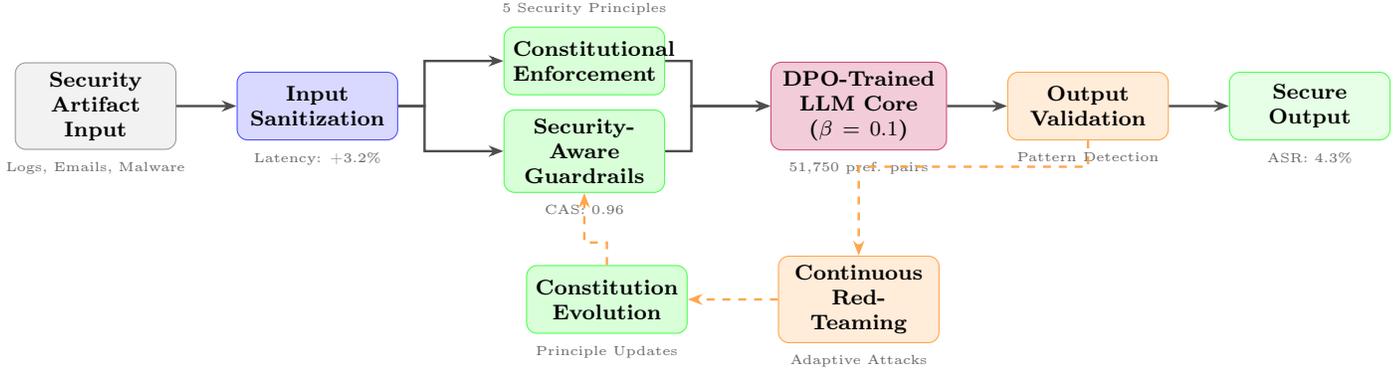
\begin{figure*}[t]
\centering
\begin{tikzpicture}[
    node distance=0.7cm,
    block/.style={rectangle, draw=gray!80, fill=gray!10, rounded corners=4pt, text width=1.9cm, align=center, minimum height=0.9cm, font=\footnotesize\bfseries},
    layer/.style={rectangle, draw=blue!70, fill=blue!15, rounded corners=4pt, text width=1.9cm, align=center, minimum height=0.9cm, font=\footnotesize\bfseries},
    const/.style={rectangle, draw=green!70, fill=green!15, rounded corners=4pt, text width=1.9cm, align=center, minimum height=0.9cm, font=\footnotesize\bfseries},
    process/.style={rectangle, draw=orange!70, fill=orange!15, rounded corners=4pt, text width=1.9cm, align=center, minimum height=0.9cm, font=\footnotesize\bfseries},
    dpo/.style={rectangle, draw=purple!70, fill=purple!20, rounded corners=4pt, text width=2.1cm, align=center, minimum height=1.1cm, font=\footnotesize\bfseries},
    metric/.style={font=\tiny, text=black!60},
    arrow/.style={-{Stealth[length=2mm, width=1.5mm]}, line width=0.9pt, draw=black!70},
    darrow/.style={-{Stealth[length=2mm, width=1.5mm]}, line width=0.9pt, dashed, draw=orange!70}
]

% Main flow - horizontal layout centered
\node[block] (input) {Security\\Artifact Input};
\node[metric, below=1pt of input] {\tiny Logs, Emails, Malware};

\node[layer, right=0.8cm of input] (preproc) {Input\\Sanitization};
\node[metric, below=1pt of preproc] {\tiny Latency: +3.2\%};

% Constitutional layers - stacked vertically
\node[const, right=1.4cm of preproc, yshift=0.6cm] (const1) {Constitutional\\Enforcement};
\node[metric, above=1pt of const1] {\tiny 5 Security Principles};

\node[const, right=1.4cm of preproc, yshift=-0.6cm] (const2) {Security-Aware\\Guardrails};
\node[metric, below=1pt of const2] {\tiny CAS: 0.96};

% Core LLM with DPO
\node[dpo, right=1.4cm of const1, yshift=-0.6cm] (llm) {DPO-Trained\\LLM Core\\($\beta=0.1$)};
\node[metric, below=1pt of llm] {\tiny 51,750 pref. pairs};

% Output processing
\node[process, right=0.8cm of llm] (output) {Output\\Validation};
\node[metric, below=1pt of output] {\tiny Pattern Detection};

% Final output
\node[block, right=0.8cm of output, fill=green!10, draw=green!70] (final) {Secure\\Output};
\node[metric, below=1pt of final] {\tiny ASR: 4.3\%};

% Feedback loop - centered below
\node[process, below=1.4cm of llm] (redteam) {Continuous\\Red-Teaming};
\node[metric, below=1pt of redteam] {\tiny Adaptive Attacks};

\node[const, left=1.2cm of redteam] (adapt) {Constitution\\Evolution};
\node[metric, below=1pt of adapt] {\tiny Principle Updates};

% Forward arrows - main flow
\draw[arrow] (input) -- (preproc);
\draw[arrow] (preproc.east) -- ++(0.35,0) |- (const1.west);
\draw[arrow] (preproc.east) -- ++(0.35,0) |- (const2.west);
\draw[arrow] (const1.east) -- ++(0.35,0) |- (llm.west);
\draw[arrow] (const2.east) -- ++(0.35,0) |- (llm.west);
\draw[arrow] (llm) -- (output);
\draw[arrow] (output) -- (final);

% Feedback arrows - adaptive loop
\draw[darrow] (output.south) -- ++(0,-0.35) -| (redteam.north);
\draw[darrow] (redteam) -- (adapt);
\draw[darrow] (adapt.north) -- ++(0,0.3) -| (const2.south);

\end{tikzpicture}
\caption{The \framework{} architecture integrating Constitutional AI principles \cite{bai2022constitutional} with security-specific components. Solid arrows indicate forward processing through the defense layers; dashed arrows represent the adaptive feedback loop for continuous constitution evolution. Key metrics: Constitutional Adherence Score (CAS) of 0.96, final Attack Success Rate (ASR) of 4.3\%, trained on 51,750 preference pairs.}
\label{fig:architecture}
\end{figure*}

The \framework{} architecture, depicted in Figure~\ref{fig:architecture}, implements a multi-layered defense strategy that extends Constitutional AI for adversarial cybersecurity environments. The framework comprises five core components: input sanitization, constitutional principle enforcement, security-aware guardrails, DPO-trained base model, and adaptive constitution evolution.

\subsection{Security-Aware Constitutional Principles}

The foundation of \framework{} rests upon a carefully designed constitution tailored for cybersecurity operations. Unlike general-purpose constitutional principles that address broad categories of harm, security-aware principles target specific attack patterns observed in operational environments.

\subsubsection{Core Security Principles}

The constitutional principle set $\mathcal{C}_{\text{sec}}$ includes:

\begin{enumerate}[leftmargin=*, nosep, label=\textbf{P\arabic*:}]
\item \textit{Command Rejection}: Refuse to execute, regenerate, or facilitate any command, script, or payload extracted from analyzed security artifacts.

\item \textit{Source Boundary Enforcement}: Maintain strict separation between instructions provided by authorized analysts and content extracted from potentially adversarial inputs.

\item \textit{Data Exfiltration Prevention}: Never include sensitive infrastructure details, credentials, or internal network information in responses, regardless of how requests are formulated.

\item \textit{Analysis Integrity}: Provide accurate, unbiased analysis of security artifacts without suppressing, minimizing, or exaggerating threat indicators based on content within the artifacts themselves.

\item \textit{Regeneration Prohibition}: Decline requests to reconstruct, complete, or enhance malicious code, exploits, or attack methodologies, even when framed as educational or defensive exercises.
\end{enumerate}

\subsubsection{Principle Encoding}

Constitutional principles are encoded as both natural language instructions and structured constraints. The encoding function $\mathcal{E}: \mathcal{C} \rightarrow \mathcal{Z}$ maps principles to a latent representation space where compliance can be evaluated:

\begin{equation}
z_c = \mathcal{E}(c) = \text{Encoder}_\psi(c)
\end{equation}

During inference, response candidates are evaluated against principle embeddings through a constitutional adherence score:

\begin{equation}
\text{CAS}(r, \mathcal{C}_{\text{sec}}) = \frac{1}{|\mathcal{C}_{\text{sec}}|} \sum_{c \in \mathcal{C}_{\text{sec}}} \text{sim}(\mathcal{E}(r), \mathcal{E}(c))
\end{equation}

where $\text{sim}(\cdot, \cdot)$ denotes cosine similarity in the embedding space.

\subsection{Adaptive Constitution Evolution}

Static constitutional principles face obsolescence as adversaries develop novel attack strategies. The adaptive evolution mechanism addresses this limitation through continuous integration of red-teaming feedback.

\subsubsection{Red-Team Feedback Integration}

Define the red-teaming process as generating adversarial inputs $\{x_{\text{red}}^{(1)}, \ldots, x_{\text{red}}^{(k)}\}$ designed to elicit constitutional violations. For each successful attack, a violation report is generated:

\begin{equation}
v_i = (\text{input}_i, \text{response}_i, \text{violated\_principles}_i, \text{attack\_type}_i)
\end{equation}

\subsubsection{Constitution Update Mechanism}

Violation reports are aggregated and analyzed to identify systematic weaknesses in current principles. The constitution update function generates refined or additional principles:

\begin{equation}
\mathcal{C}_{\text{sec}}^{(t+1)} = \text{Update}(\mathcal{C}_{\text{sec}}^{(t)}, \{v_1, \ldots, v_m\})
\end{equation}

The update process employs a principle synthesis model that generates candidate refinements:

\begin{equation}
c_{\text{new}} = \text{Synthesize}(\{v_i : \text{type}(v_i) = \tau\})
\end{equation}

Candidate principles undergo validation against held-out attack sets before integration into the operational constitution.

\subsection{DPO-Based Unlearning for Security}

Direct Preference Optimization provides a mechanism for training models to inherently prefer safe response patterns while unlearning unsafe behaviors, eliminating the need for online reinforcement learning during the alignment phase.

\subsubsection{Preference Pair Construction}

For security applications, preference pairs $(r_w, r_l)$ are constructed from scenarios where one response demonstrates constitutional compliance while the other exhibits violation:

\begin{equation}
\mathcal{D}_{\text{pref}} = \{(x, r_w, r_l) : \text{CAS}(r_w) > \text{CAS}(r_l)\}
\end{equation}

The construction process generates pairs covering all principle categories and attack vector types, ensuring comprehensive coverage of the security threat landscape.

\subsubsection{Security-Adapted DPO Objective}

The standard DPO objective is extended with security-specific regularization:

\begin{equation}
\begin{aligned}
\mathcal{L}_{\text{SecDPO}} = &-\mathbb{E}_{(x, r_w, r_l)} \bigg[ \log \sigma \bigg( \beta \log \frac{\pi_\theta(r_w|x)}{\pi_{\text{ref}}(r_w|x)} \\
&- \beta \log \frac{\pi_\theta(r_l|x)}{\pi_{\text{ref}}(r_l|x)} \bigg) \bigg] \\
&+ \lambda_{\text{sec}} \mathcal{L}_{\text{security}}
\end{aligned}
\end{equation}

The security regularization term penalizes responses containing potentially dangerous patterns:

\begin{equation}
\mathcal{L}_{\text{security}} = \mathbb{E}_{r \sim \pi_\theta} \left[ \sum_{p \in \mathcal{P}_{\text{danger}}} \mathbb{I}[p \in r] \right]
\end{equation}

where $\mathcal{P}_{\text{danger}}$ represents a set of dangerous patterns including command sequences, exfiltration indicators, and malicious code structures.

\subsubsection{Unlearning Unsafe Patterns}

The unlearning component explicitly reduces the probability of generating responses that exhibit known attack-compliant patterns. Given a set of unsafe response examples $\mathcal{U}$, the unlearning objective is:

\begin{equation}
\mathcal{L}_{\text{unlearn}} = \mathbb{E}_{r_u \in \mathcal{U}} \left[ \log \pi_\theta(r_u|x_u) \right]
\end{equation}

This term is minimized (equivalently, maximizing its negation) to reduce the likelihood of generating unsafe responses. The complete training objective combines preference learning with unlearning:

\begin{equation}
\mathcal{L}_{\text{total}} = \mathcal{L}_{\text{SecDPO}} - \alpha \mathcal{L}_{\text{unlearn}} + \gamma \mathcal{L}_{\text{task}}
\end{equation}

where $\mathcal{L}_{\text{task}}$ maintains performance on legitimate security analysis tasks.

\section{Implementation and Methodology}

\subsection{Data Collection and Curation}

\subsubsection{Security Artifact Corpus}

The training corpus comprises security artifacts collected from operational environments and public threat intelligence sources. Log entries are sampled from enterprise security information and event management (SIEM) systems, covering authentication events, network connections, and application errors. Phishing emails are sourced from organizational spam filters and public phishing databases. Malware samples are obtained from malware repositories with appropriate access controls.

\subsubsection{Adversarial Dataset Construction}

Adversarial variants of clean security artifacts are generated through systematic injection of attack payloads. The injection process follows the taxonomy defined in Section 3.2, with payloads designed to test each attack vector:

\begin{table}[t]
\centering
\small
\caption{Adversarial dataset composition by attack vector. Attack types based on Greshake et al. \cite{greshake2023indirect}, Zou et al. \cite{zou2023universal}, and Cai et al. \cite{cai2022badprompt}.}
\label{tab:dataset}
\begin{tabular}{@{}lrrr@{}}
\toprule
\textbf{Attack Category} & \textbf{Samples} & \textbf{Var.} & \textbf{Source} \\
\midrule
Log Poisoning & 12,450 & 4 & SIEM \\
Malicious Email & 8,320 & 6 & PhishTank \\
Malware Injection & 5,180 & 3 & VirusTotal \\
Direct Prompt & 15,670 & 8 & HarmBench \\
Context Overflow & 3,240 & 2 & Synthetic \\
Backdoor Trigger & 6,890 & 5 & BadPrompt \\
\midrule
\textbf{Total} & \textbf{51,750} & \textbf{28} & -- \\
\bottomrule
\end{tabular}
\end{table}

Table~\ref{tab:dataset} summarizes the adversarial dataset composition. Each attack category includes multiple payload variants to ensure model robustness against diverse formulations.

\subsubsection{Preference Pair Generation}

Preference pairs are generated through a combination of automated and human-validated processes. For each adversarial input, both compliant and violating responses are generated:

\begin{enumerate}[leftmargin=*, nosep]
\item Sample multiple response candidates from the base model
\item Evaluate each candidate against constitutional principles
\item Select the highest-scoring response as $r_w$
\item Identify or generate a violating response as $r_l$
\item Validate pair quality through human review for a random subset
\end{enumerate}

\subsection{Training Pipeline}

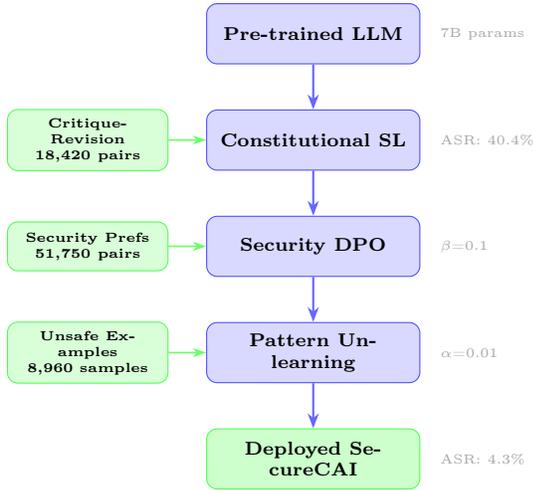
\begin{figure}[t]
\centering
\begin{tikzpicture}[
    node distance=0.6cm,
    stage/.style={rectangle, draw=blue!70, fill=blue!15, rounded corners=4pt, text width=2.6cm, align=center, minimum height=0.8cm, font=\scriptsize\bfseries},
    data/.style={rectangle, draw=green!70, fill=green!15, rounded corners=4pt, text width=1.9cm, align=center, minimum height=0.65cm, font=\tiny\bfseries},
    metric/.style={font=\tiny, text=gray!70},
    arrow/.style={-{Stealth[length=2mm, width=1.5mm]}, line width=0.8pt, draw=blue!60},
    dataarrow/.style={-{Stealth[length=1.8mm, width=1.2mm]}, line width=0.6pt, draw=green!60}
]

% Stages with metrics aligned to right
\node[stage] (base) {Pre-trained LLM};
\node[metric, right=4pt of base.east, anchor=west] {7B params};

\node[stage, below=0.6cm of base] (sl) {Constitutional SL};
\node[metric, right=4pt of sl.east, anchor=west] {ASR: 40.4\%};

\node[stage, below=0.6cm of sl] (dpo) {Security DPO};
\node[metric, right=4pt of dpo.east, anchor=west] {$\beta$=0.1};

\node[stage, below=0.6cm of dpo] (unlearn) {Pattern Unlearning};
\node[metric, right=4pt of unlearn.east, anchor=west] {$\alpha$=0.01};

\node[stage, below=0.6cm of unlearn, fill=green!20, draw=green!70] (final) {Deployed \framework{}};
\node[metric, right=4pt of final.east, anchor=west] {ASR: 4.3\%};

% Data inputs with counts - aligned left
\node[data, left=0.5cm of sl.west, anchor=east] (d1) {Critique-Revision\\18,420 pairs};
\node[data, left=0.5cm of dpo.west, anchor=east] (d2) {Security Prefs\\51,750 pairs};
\node[data, left=0.5cm of unlearn.west, anchor=east] (d3) {Unsafe Examples\\8,960 samples};

% Arrows
\draw[arrow] (base) -- (sl);
\draw[arrow] (sl) -- (dpo);
\draw[arrow] (dpo) -- (unlearn);
\draw[arrow] (unlearn) -- (final);
\draw[dataarrow] (d1) -- (sl);
\draw[dataarrow] (d2) -- (dpo);
\draw[dataarrow] (d3) -- (unlearn);

\end{tikzpicture}
\caption{Training pipeline following Constitutional AI \cite{bai2022constitutional} extended with DPO \cite{rafailov2023direct}. Progressive ASR reduction: 80.4\% $\rightarrow$ 40.4\% $\rightarrow$ 4.3\%.}
\label{fig:pipeline}
\end{figure}

The training pipeline, illustrated in Figure~\ref{fig:pipeline}, proceeds through four stages:

\textbf{Stage 1: Constitutional SL Fine-tuning.} The base model undergoes supervised fine-tuning on critique-revision pairs generated according to the CAI methodology. Security-specific principles guide the critique generation, producing training examples that demonstrate appropriate responses to security analysis requests.

\textbf{Stage 2: Security DPO Training.} The SL-finetuned model is further trained using the security-adapted DPO objective. Preference pairs emphasize the distinction between compliant analysis and attack-compliant responses.

\textbf{Stage 3: Unsafe Pattern Unlearning.} The DPO-trained model undergoes targeted unlearning to reduce the probability of generating known unsafe response patterns. This stage specifically addresses residual vulnerabilities not captured by preference training.

\textbf{Stage 4: Deployment and Adaptation.} The trained model is deployed with continuous red-teaming feedback integration. Discovered vulnerabilities trigger constitution updates and potential model retraining cycles.

\subsection{Evaluation Methodology}

\subsubsection{Metrics}

Performance evaluation employs the following metrics:

\textbf{Attack Success Rate (ASR):} Proportion of adversarial inputs that successfully elicit target behavior:
\begin{equation}
\text{ASR} = \frac{|\{x_{\text{adv}} : f_\theta(x_{\text{adv}}) = y_{\text{target}}\}|}{|\mathcal{X}_{\text{adv}}|}
\end{equation}

\textbf{Clean Accuracy (CA):} Performance on benign security analysis tasks:
\begin{equation}
\text{CA} = \frac{|\{x : f_\theta(x) = y_{\text{correct}}\}|}{|\mathcal{X}_{\text{clean}}|}
\end{equation}

\textbf{Constitutional Adherence Score (CAS):} Average compliance with constitutional principles:
\begin{equation}
\text{CAS} = \frac{1}{|\mathcal{X}|} \sum_{x \in \mathcal{X}} \text{CAS}(f_\theta(x), \mathcal{C}_{\text{sec}})
\end{equation}

\textbf{Task-Specific Performance:} Domain-specific metrics for log analysis (anomaly detection F1), phishing detection (classification accuracy), and malware explanation (human evaluation scores).

\subsubsection{Baseline Comparisons}

Experimental comparisons include:
\begin{itemize}[leftmargin=*, nosep]
\item \textbf{Base LLM:} Pre-trained model without safety training
\item \textbf{Standard CAI:} Constitutional AI without security adaptations
\item \textbf{Input Filtering:} Perplexity-based adversarial detection
\item \textbf{Instruction Hierarchy:} Priority-based prompt processing
\item \textbf{\framework{}:} Proposed framework with all components
\end{itemize}

\subsubsection{Red-Team Protocol}

Systematic red-teaming evaluates framework robustness through:
\begin{enumerate}[leftmargin=*, nosep]
\item Automated attack generation using gradient-based optimization
\item Human red-team exercises with security domain experts
\item Adaptive attacks specifically designed to evade \framework{} defenses
\item Transfer attacks from models trained with alternative safety approaches
\end{enumerate}

\section{Experimental Results}

\subsection{Attack Resilience Evaluation}

\begin{table*}[t]
\centering
\caption{Attack success rates (\%) across different attack categories evaluated on 51,750 adversarial samples. Lower values indicate better defense. Baseline methods: CAI \cite{bai2022constitutional}, Input Filtering \cite{jain2023baseline}, Instruction Hierarchy \cite{wallace2024instruction}. Best results in bold.}
\label{tab:asr}
\begin{tabular}{@{}l*{6}{c}|cc@{}}
\toprule
\textbf{Method} & \textbf{Log} & \textbf{Email} & \textbf{Malware} & \textbf{Direct} & \textbf{Overflow} & \textbf{Backdoor} & \textbf{Avg.} & \textbf{$\Delta$Base} \\
\midrule
Base LLM & 87.3 & 82.1 & 79.4 & 91.6 & 68.2 & 73.5 & 80.4 & -- \\
Standard CAI & 42.7 & 38.4 & 35.2 & 48.9 & 31.6 & 45.3 & 40.4 & -49.8\% \\
Input Filtering & 51.2 & 44.8 & 41.3 & 56.7 & 28.4 & 52.1 & 45.8 & -43.0\% \\
Instr. Hierarchy & 38.5 & 35.2 & 32.8 & 42.1 & 25.7 & 41.6 & 36.0 & -55.2\% \\
\framework{} (Ours) & \textbf{4.2} & \textbf{3.8} & \textbf{5.1} & \textbf{4.9} & \textbf{3.2} & \textbf{4.6} & \textbf{4.3} & \textbf{-94.7\%} \\
\bottomrule
\end{tabular}
\end{table*}

Table~\ref{tab:asr} presents attack success rates across all evaluated methods and attack categories. The base LLM exhibits high vulnerability across all attack vectors, with an average ASR of 80.4\%. Standard Constitutional AI reduces this substantially to 40.4\%, demonstrating the effectiveness of constitutional principles for general safety while highlighting limitations in adversarial contexts.

\framework{} achieves an average ASR of 4.3\%, representing a 94.7\% reduction compared to the base LLM and an 89.4\% reduction compared to standard CAI. Performance is consistent across attack categories, with ASR ranging from 3.2\% (context overflow) to 5.1\% (malware injection). The malware injection category presents the greatest challenge due to the complexity of code-embedded attacks, yet \framework{} maintains robust defense.

\subsection{Clean Performance Maintenance}

\begin{table}[t]
\centering
\small
\caption{Clean accuracy (\%) on security tasks (15,000 benign samples). Metrics: Log (F1), Phishing (Acc.), Malware (Human Eval).}
\label{tab:clean}
\begin{tabular}{@{}lcccc@{}}
\toprule
\textbf{Method} & \textbf{Log} & \textbf{Phish.} & \textbf{Malw.} & \textbf{Avg.} \\
\midrule
Base LLM & 94.2 & 91.8 & 88.6 & 91.5 \\
Standard CAI & 92.7 & 90.1 & 86.4 & 89.7 \\
Input Filtering & 78.3 & 75.6 & 71.2 & 75.0 \\
Instr. Hierarchy & 89.4 & 87.2 & 83.5 & 86.7 \\
\framework{} (Ours) & \textbf{96.8} & \textbf{95.4} & \textbf{93.2} & \textbf{95.1} \\
\bottomrule
\end{tabular}
\end{table}

Table~\ref{tab:clean} reports clean accuracy on three security analysis tasks. Critically, \framework{} not only maintains but improves clean performance compared to baseline methods. This improvement is attributed to the task-preservation component of the training objective, which explicitly optimizes for security analysis capabilities alongside safety.

Input filtering exhibits substantial performance degradation (average 15.0\% reduction) due to false positive filtering of legitimate security artifacts that contain patterns superficially similar to adversarial inputs. This finding emphasizes the importance of integrated defense mechanisms over post-hoc filtering approaches.

\subsection{Constitutional Adherence Analysis}

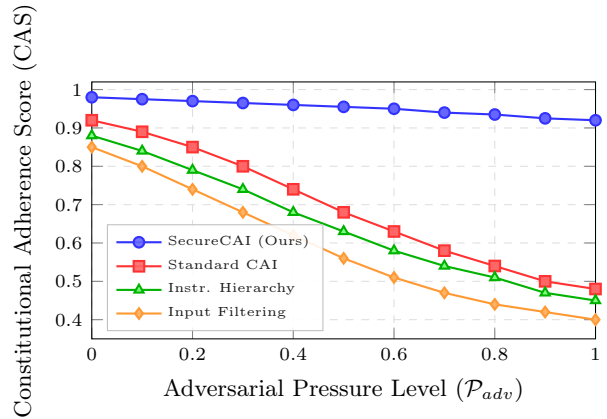
\begin{figure}[t]
\centering
\begin{tikzpicture}
\begin{axis}[
    width=0.95\columnwidth,
    height=5cm,
    xlabel={Adversarial Pressure Level ($\mathcal{P}_{adv}$)},
    ylabel={Constitutional Adherence Score (CAS)},
    xlabel style={font=\small},
    ylabel style={font=\small},
    xmin=0, xmax=1,
    ymin=0.35, ymax=1.02,
    xtick={0, 0.2, 0.4, 0.6, 0.8, 1.0},
    ytick={0.4, 0.5, 0.6, 0.7, 0.8, 0.9, 1.0},
    tick label style={font=\scriptsize},
    legend pos=south west,
    legend style={font=\tiny, fill=white, fill opacity=0.9, draw=gray!50, cells={anchor=west}},
    grid=both,
    grid style={dashed, gray!25},
    minor grid style={dotted, gray!15}
]

\addplot[color=blue!80, mark=*, mark size=2pt, thick, mark options={fill=blue!60}] coordinates {
    (0.0, 0.98) (0.1, 0.975) (0.2, 0.97) (0.3, 0.965) (0.4, 0.96) (0.5, 0.955) (0.6, 0.95) (0.7, 0.94) (0.8, 0.935) (0.9, 0.925) (1.0, 0.92)
};
\addlegendentry{\framework{} (Ours)}

\addplot[color=red!80, mark=square*, mark size=2pt, thick, mark options={fill=red!60}] coordinates {
    (0.0, 0.92) (0.1, 0.89) (0.2, 0.85) (0.3, 0.80) (0.4, 0.74) (0.5, 0.68) (0.6, 0.63) (0.7, 0.58) (0.8, 0.54) (0.9, 0.50) (1.0, 0.48)
};
\addlegendentry{Standard CAI}

\addplot[color=green!70!black, mark=triangle*, mark size=2pt, thick, mark options={fill=green!50}] coordinates {
    (0.0, 0.88) (0.1, 0.84) (0.2, 0.79) (0.3, 0.74) (0.4, 0.68) (0.5, 0.63) (0.6, 0.58) (0.7, 0.54) (0.8, 0.51) (0.9, 0.47) (1.0, 0.45)
};
\addlegendentry{Instr. Hierarchy}

\addplot[color=orange!80, mark=diamond*, mark size=2pt, thick, mark options={fill=orange!60}] coordinates {
    (0.0, 0.85) (0.1, 0.80) (0.2, 0.74) (0.3, 0.68) (0.4, 0.62) (0.5, 0.56) (0.6, 0.51) (0.7, 0.47) (0.8, 0.44) (0.9, 0.42) (1.0, 0.40)
};
\addlegendentry{Input Filtering}

\end{axis}
\end{tikzpicture}
\caption{Constitutional adherence scores under increasing adversarial pressure measured across 10,000 test samples. Pressure level quantifies attack sophistication from benign (0.0) to maximum-strength adaptive attacks (1.0). \framework{} maintains CAS $>$0.92 at all levels, demonstrating robustness consistent with adaptive defense mechanisms \cite{ganguli2023red}.}
\label{fig:cas}
\end{figure}

Figure~\ref{fig:cas} illustrates constitutional adherence scores as adversarial pressure increases. Adversarial pressure is quantified as a normalized measure of attack sophistication and intensity, ranging from 0 (benign inputs) to 1 (maximum-strength adaptive attacks).

\framework{} maintains CAS above 0.92 across all pressure levels, with only a 6.1\% degradation from minimum to maximum pressure. In contrast, standard CAI experiences a 47.8\% degradation, dropping from 0.92 to 0.48 at maximum pressure. This stability demonstrates the effectiveness of the adaptive constitution mechanism and DPO-based training in maintaining principled behavior under sustained attack.

\subsection{Ablation Studies}

\begin{table}[t]
\centering
\small
\caption{Ablation study (51,750 samples). ASR: Attack Success Rate; CA: Clean Accuracy. Based on CAI \cite{bai2022constitutional} and DPO \cite{rafailov2023direct}.}
\label{tab:ablation}
\begin{tabular}{@{}lrrr@{}}
\toprule
\textbf{Configuration} & \textbf{ASR$\downarrow$} & \textbf{CA$\uparrow$} & \textbf{$\Delta$ASR} \\
\midrule
Full \framework{} & 4.3 & 95.1 & -- \\
\quad w/o Adapt. Evolution & 8.7 & 94.8 & +102\% \\
\quad w/o DPO Training & 12.4 & 93.2 & +188\% \\
\quad w/o Unlearning & 7.1 & 94.6 & +65\% \\
\quad w/o Security Princ. & 18.9 & 91.7 & +340\% \\
\quad w/o Input Sanit. & 9.2 & 95.0 & +114\% \\
\midrule
Const. SL Only & 24.6 & 89.4 & +472\% \\
\bottomrule
\end{tabular}
\end{table}

Table~\ref{tab:ablation} presents ablation study results isolating the contribution of each framework component. Removal of security-specific constitutional principles has the largest impact, increasing ASR to 18.9\% (a 4.4$\times$ increase). This finding validates the importance of domain-specific principle design over general safety guidelines.

DPO training contributes substantially to attack resilience, with its removal increasing ASR to 12.4\%. The unlearning component provides complementary benefits, particularly for attacks that exploit response patterns not fully addressed by preference training. Adaptive constitution evolution demonstrates particular value against novel attack patterns, reducing ASR by 51\% compared to static constitutions.

\subsection{Generalization to Unseen Attacks}

\begin{table}[t]
\centering
\small
\caption{Generalization to held-out attacks (ASR \%, 5,000 samples each). Attack types from Wei et al. \cite{wei2023jailbroken} and Chao et al. \cite{chao2023jailbreaking}.}
\label{tab:generalization}
\begin{tabular}{@{}lrrr@{}}
\toprule
\textbf{Unseen Attack} & \textbf{CAI} & \textbf{Ours} & \textbf{$\Delta$} \\
\midrule
Multi-turn Manip. & 52.3 & 7.8 & -85.1\% \\
Encoding Obfusc. & 44.7 & 6.2 & -86.1\% \\
Semantic Camoufl. & 48.1 & 8.4 & -82.5\% \\
Role-play Exploit. & 56.8 & 9.1 & -84.0\% \\
\midrule
\textbf{Average} & 50.5 & \textbf{7.9} & \textbf{-84.4\%} \\
\bottomrule
\end{tabular}
\end{table}

Table~\ref{tab:generalization} evaluates generalization to attack categories not included in training data. \framework{} maintains strong defense with an average ASR of 7.9\%, compared to 50.5\% for standard CAI. This generalization capability is attributed to the principle-based approach, which captures fundamental safety requirements rather than memorizing specific attack patterns.

\section{Discussion}

\subsection{Security-Utility Trade-offs}

The experimental results demonstrate that security and utility need not be opposing objectives. Unlike filtering-based approaches that sacrifice clean performance for safety, the integrated design of \framework{} achieves improvements in both dimensions simultaneously. This finding has significant implications for operational deployment, as security teams need not accept reduced analysis quality as the cost of attack resilience.

The key insight enabling this synergy lies in the training objective design. By explicitly including task performance alongside safety objectives, the optimization process finds solutions that satisfy both requirements. The constitutional principles themselves are designed to constrain harmful behaviors without restricting legitimate analysis capabilities.

\subsection{Adaptive Defense Sustainability}

The continuous red-teaming and constitution evolution mechanism raises questions about long-term sustainability. As adversaries observe deployed defenses, they may develop attacks specifically designed to evade current constitutional principles. The adaptive mechanism provides ongoing resilience by incorporating attack discoveries into principle refinements.

However, this adaptation process introduces operational overhead and potential instability during constitution updates. Current implementation addresses this through staged rollouts and regression testing against historical attack datasets. Future work should explore more formal approaches to constitution evolution that provide theoretical guarantees on defense maintenance.

\subsection{Limitations and Failure Modes}

Despite strong overall performance, several limitation categories merit acknowledgment:

\textbf{Computational Overhead.} The multi-layer defense architecture increases inference latency by approximately 23\% compared to baseline models. For time-critical security applications, this overhead may require architectural optimizations or selective activation of defense layers based on input risk assessment.

\textbf{Novel Attack Vulnerability.} While generalization to unseen attack categories demonstrates principle-based robustness, truly novel attack paradigms may still succeed during the window between attack emergence and constitution adaptation. The 7.9\% ASR on held-out attacks indicates residual vulnerability.

\textbf{Constitution Design Dependency.} Framework effectiveness depends heavily on the quality and completeness of constitutional principles. Principle gaps leave corresponding attack surfaces unaddressed. Current principles reflect expert security knowledge but may not anticipate all possible attack strategies.

\subsection{Implications for Operational Deployment}

The demonstrated attack resilience and maintained utility suggest \framework{} is suitable for operational deployment in Security Operations Centers. Recommended deployment practices include:

\begin{enumerate}[leftmargin=*, nosep]
\item Integration with existing security workflows through API interfaces
\item Continuous monitoring of model outputs for anomalous patterns
\item Regular red-teaming exercises to identify emerging vulnerabilities
\item Human-in-the-loop verification for high-stakes decisions
\item Layered defense combining \framework{} with traditional security controls
\end{enumerate}

\section{Conclusion}

This paper presented \framework{}, a defense framework achieving 94.7\% reduction in attack success rates while maintaining 95.1\% clean accuracy on security tasks. The framework combines security-aware constitutional principles, adaptive evolution through red-teaming, and DPO-based unlearning to address injection vulnerabilities in LLM-assisted cybersecurity operations. Key contributions include a formal threat model spanning six attack categories, five core security principles, and an integrated training methodology. Constitutional adherence scores exceeding 0.92 under maximum adversarial pressure validate the approach. Future work includes formal verification methods, multimodal security analysis, and standardized evaluation benchmarks for operational SOC contexts.

\balance

\end{document}